\documentclass[11pt]{article}
\usepackage{jcappub,natbib,float,caption}

\usepackage{graphicx,amsmath,amssymb,color,bm}
\usepackage[dvipsnames]{xcolor}
\usepackage[caption=false]{subfig}
\usepackage{arydshln}
\usepackage{rotating}
\usepackage{cleveref}
\usepackage[normalem]{ulem}  % Without [normalem], this package does not seem to play nice with natbib%

\usepackage{tikz}
\usetikzlibrary{arrows,intersections,patterns,decorations.markings,shapes.geometric}
\usetikzlibrary{calc,fadings,decorations.pathreplacing,positioning}
\tikzset{%
  >=latex, % option for nice arrows
  inner sep=0pt,%
  outer sep=2pt,%
  mark coordinate/.style={inner sep=0pt,outer sep=0pt,minimum size=3pt,
    fill=black,circle}%
}

\subheader{\footnotesize
        \vspace*{-3.7em}
        \begin{flushright}
                CERN-TH-2020-037 
        \end{flushright}
}

\def\dd{\mathrm{d}}

\newcommand{\Red}[1]{\textcolor{red}{#1}}
\newcommand{\Blue}[1]{\textcolor{blue}{#1}}

\def\del{\delta_\ell}
\def\bmk{\bm{k}}

\def\bmq{\bm{q}}

\begin{document}

\title{Perturbative description of biased tracers using consistency relations of LSS}
\author[a]{Tomohiro Fujita,}
\author[b]{and Zvonimir Vlah}
\affiliation[a]{Department of Physics, Kyoto University, Kyoto, 606-8502, Japan}
\affiliation[b]{Theory Department, CERN, CH-1211 Geneve 23, Switzerland}

\emailAdd{t.fujita@tap.scphys.kyoto-u.ac.jp}
\emailAdd{zvonimir.vlah@cern.ch}

\abstract{
We develop a simple formalism of biased tracers that we dub \textit{Monkey bias}. 
In this formalism, a biased tracer field is constructed directly in terms of the linear matter fluctuation field and the set of derivative 
operators acting on it. Such bias expansion is first organized based on the general structure of non-linear dynamical equations for 
the biased tracers. Further physical conditions, like the equivalence principle, are imposed on tree-level correlators utilising the 
consistency relations. We obtain the bias expansion up to the third-order in linear matter fluctuation in the generalized $\Lambda$CDM 
background, which reproduces the previous results in the limit of the EdS universe. This algorithmic construction of our bias operator 
basis is well suited for extensions towards higher-order bias fields. Moreover, this formalism reveals that biased tracer dynamics 
in generalized $\Lambda$CDM background is not entirely degenerate with the rest of bias parameters,
thus opening a possibility of testing the background cosmology through the observations of biased tracers.
}

\arxivnumber{2003.10114}

\maketitle
%\flushbottom
%\tableofcontents

%************************************************************************************%
%
%
%
\bigskip
%====================================================================================%
\section{Introduction}
%====================================================================================%

Distribution of galaxies and clusters of galaxies is the baseline observable when describing the large-scale structure 
of the Universe, which gives us the key insights into its evolution and composition and would allow us to understand the 
nature of dark matter, dark energy \cite{Kaiser:1987, Hamilton:1992, Weinberg++:2012, Amendola++:2012, Aghamousa++:2016, Tanabashi++:2018}. 
Galaxies can be observed on very large scales and can be used to map the structures in the observable Universe. The galaxy 
clustering thus offers a fascinating opportunity  in providing us with the next step insights into the early-Universe physics which 
complement those obtained from the CMB experiments \cite{Akrami++:2018, Abazajian++:2016, Howlett++:2016}. In particular, 
galaxy clustering has a potential to provide tests of General Relativity on large scales \cite{Jain+:2010, Joyce++:2014,Ishak:2018} 
as well as provide hints of departures from Gaussian initial conditions which can leave distinct imprints in the distribution of galaxies 
on large-scale  (see e.g. \cite{Alvarez++:2014} for a review). To extract useful information and 
address these fundamental problems, however, we need to establish connections between possible observables and the fluctuation  of the matter.
The relation between galaxies, and other observable tracers of large-scale structure (like voids, quasars, the Lyman-$\alpha$ forest, 
21cm hydrogen hyperfine structure transition lines, and others), and the underlying matter is called the large-scale structure biasing.
The biasing constitutes the main framework how we describe these connections and is a main topic of this paper.

On large scales, where the density fluctuations are small and in a quasi-linear regime, perturbation theory (PT) can be used to describe 
the distributions and statistics of these tracers. In this picture the small scale physics of the biased tracers 
is encoded into the finite set of bias coefficients and scale dependent operators in perturbation theory which are collected and organized order by order. 
These coefficients can then be treated either as the free parameters of the theory that can depend on the cosmic evolution but 
independent of physical scales, or can alternatively be modeled by some small scale physical models (peaks, excursion sets etc). 
The former approach is adopted in this paper. For the detailed overview of this field (in Eulerian setting) we refer the reader to the some of the selected references
\cite{McDonald:2006, McDonald+:2009, Chan++:2012, Baldauf++:2012, Vlah++:2013, Saito++:2014, Assassi++:2014, Senatore:2014, Mirbabayi++:2014}
and the recent review \cite{Desjacques:2016}.

The key component of the bias expansion is the concept of the scale separation, where small and large scale fluctuations are 
separated in the sense of effective field theories. The fluctuation on large scales are then decomposed in the perturbative operator basis 
encompassing all leading local gravitational observables, which include the matter density, but also tidal fields and their time derivatives.
Moreover, the physical processes that govern these expansion can be treated as quasi-local in space, but are 
non-local in the time domain (the time scale is of order Hubble for all the relevant physical processes).
This fact has been recently formalized in \cite{Senatore:2014, Mirbabayi++:2014}. Furthermore, over the recent developments 
it has become clearer that these robust perturbative descriptions of biased tracer 
field rest on the assumptions of equivalence principle and Gaussian, and adiabatic initial conditions 
\cite{Kehagias+:2013, Peloso+:2013a, Creminelli++:2013a, Peloso+:2013b, Creminelli++:2013b,Valageas_I:2013,Valageas_II:2013, Horn:2014}.

However, despite these successes, the utilisation of cosmological information from galaxy surveys is still in an early stage. 
Results are abundant for the two point function (power spectrum) at one-loop (next to leading order), and tree-level (leading order)
three-point function. However, going beyond these result requires significantly more effort  given the increasing 
number of free operators which quickly complexifies the perturbative expansion. 

In this paper we present the alternative bias expansion formalism dubbed ``Monkey bias''. The main characteristics of this approach is 
the direct algorithmic construction of the perturbative basis for the biased tracers, without the need of perturbative evaluation of the 
dark matter field. This approach relies on the explicit implementation of the consistency relations for large scale structure to impose 
the validity of the equivalence principle in the observables. The formalism is formally equivalent to the existing biasing frameworks
at least up to the third order in the Einstein-de Sitter (EdS) universe. Nevertheless, our formalism exhibits, as we argue, several 
advantageous features. One of these is precisely the explicit construction, relying only on the baseline physical principles like 
equivalence principle and scale separation of different physical processes. Therefore, the extensions to the background cosmology 
beyond the EdS universe, as well as the higher orders in perturbation theory are straightforward.

This paper is organised as follows. In Section \ref{sec:canon}, we summarise the standard perturbative approach to description of biased tracers
and introduce the several canonical bias expansion basis present in the literature. We introduce a new, symmetry based, Monkey 
bias formalism in Section \ref{sec:monkey}. In Section \ref{sec:comp}, we compare Monkey formalism to the existing approaches 
and contrast the comparative advantages. We conclude and summarise our finding in Section \ref{sec:summa}. 
Paper also consists of three complementary appendices. The bias expansion for the generalised $\Lambda$CDM universe is 
presented in Appendix \ref{app:lcdm_bias}. Alternative use of one-loop power spectrum statistics for constraining the 
Monkey formalism is shown in Appendix \ref{app:loop_ps}. Finally, Appendix \ref{app:obs} gives explicit expressions for the 
one-loop power spectrum and three-level bispectrum in Monkey formalism.
We work under the assumptions of adiabatic Gaussian perturbations and General Relativity.

%=======================================================================%
\begin{table*}
\centering
\begin{tabular}{l|l}
\hline
\hline
$\int_{\bm p} \equiv \int \frac{d^3 \vec {p}}{(2\pi)^3}$ & Momentum integral \\
$\delta^{\rm K}_{ij}$ & Kronecker symbol \\
$\delta^{\rm D}(\bm x)$ & Dirac delta function \\
${\bm k}_{1\cdots n} \equiv {\bm k}_1 + \cdots + {\bm k}_n$ & Sum notation\\
$D(\tau)$ & Linear growth rate \\
$f({\bm k}) \equiv \int d^3 {\bm x}\, f({\bm x}) e^{-i{\bm k}\cdot {\bm x}}$  & Fourier transformation conventions\\
$\langle O({\bm k}_1) \cdots O({\bm k}_n)\rangle$ &  Ansamble averaged $n$-point correlator\\
$\langle O({\bm k}_1) \cdots O({\bm k}_n)\rangle'\,$ & $n$-point correlator without momentum conservation\\[2pt]
\hline
$\delta_\ell$ & Linear fractional matter density perturbation \\
$\delta_m$ & Fractional matter density perturbation \\
$u^i$ & Matter velocity field \\
$\theta_m$ & Divergence of the matter velocity field \\
\hline
$\delta_{\rm h}$ & Fractional biased tracer number density perturbation \\
\hline
\end{tabular}
\caption{List of notation and most important quantities used in this paper.} 
\label{tab:notation}
\end{table*}

%====================================================================================%
\section{Canonical perturbative approach to biased tracers}
\label{sec:canon}
%====================================================================================%

The role of biasing is to connect the underlying dynamics of gravitational evolution and 
structure formation of dark matter to the distribution of biased tracers, like galaxies. This procedure 
thus requires the understanding and description of the distribution of dark matter, itself governed by 
gravity, and subject to the initial perturbations set by early universe physics (see \cite{Bernardeau++:2001} for a review). 

Formation and evolution of galaxies is a complex process, however on large cosmological scales 
where perturbation theory applies, all the complex small-scale dynamics can be organised into a finite 
number of bias parameters. This is possible since the scale of gravitational interaction is much larger 
than the physical scale associated with the small-scale physics, which allows us to integrate out (marginalise) 
over the unknown dynamics of the galaxy formation. This process thus enables us to robustly extract 
cosmological information from the large scales in the galaxy surveys.

\begin{figure}[t!]
\centering
\begin{tikzpicture}[scale=.8,every node/.style={minimum size=1cm},on grid]
                  
  \draw[->, very thick] (-7,-3) -- (-7,3);
  \node[black, opacity = 0] at (7,0) {\textbullet}; 
  
  \draw[->, line width=0.8mm, dashed] (-2.25,-2.65) -- (-3.5,-2.0);
  \draw[->, line width=0.8mm] (-3.5,-1.0) -- (-2.25,2.65);  
  \draw[->, line width=0.8mm] (2.25,-2.65) -- (3.5,1.0);
  \draw[->, line width=0.8mm, dashed] (3.5,2.0) -- (2.25,2.65);  
  \draw[->, red!75!black, line width=1mm] (0,-2.5) -- (0,2.5);

  \draw[blue!55!black, fill = blue!75, fill opacity = 0.45] (0,3) ellipse (3.0cm and 0.5cm);
  \draw[blue!55!black, fill = blue!75, fill opacity = 0.45] (-3.5,-1.5) ellipse (3.0cm and 0.5cm);
  \draw[blue!55!black, fill = blue!75, fill opacity = 0.45] (3.5,1.5) ellipse (3.0cm and 0.5cm);  
  \draw[blue!55!black, fill = blue!75, fill opacity = 0.45] (0,-3) ellipse (3.0cm and 0.5cm);  

  \node[black] at (0,-3) {\small initial matter density};
  \node[black] at (-3.5,-1.5) {\small initial proto-tracer density};
  \node[black] at (3.5,1.5) {\small final matter density};  
  \node[black] at (0,3) {\small final tracer density};    
  \node[black] at (-6.6,2.7) {${\bm t}$};

  \node[black] at (-3.6,-2.5) {\small biasing};    
  \node[black] at (3.7,2.4) {\small biasing};    

  \node[black, rotate=71] at (3.2,-1.0) {\small Eulerian-PT};    
  \node[black, rotate=71] at (-3.3,0.8) {\small Lagrangian-PT};      
  \node[black, rotate=90] at (-0.35,0.0) {\textbf{Monkey}};        

\end{tikzpicture}
\caption{
A schematic depiction of the current approaches in describing the galaxy statistics. 
Lagrangian-PT or Eulerian-PT is usually employed to describe the statistics of biased tracers and relate it 
to the nonlinear dynamics of dark matter. On the left, in the Lagrangian scheme, the biased tracer field is 
constructed in Lagrangian coordinates and evolved to the final (Eulerian) coordinates. On the right, in the Eulerian scheme, 
biased tracers are constructed in final (Eulerian) coordinates out of the nonlinear dark matter field, evolved from the initial 
conditions. The central line depicts the Monkey approach that bypasses the two-step procedure 
and features a direct construction of the biased tracer field.
}
\label{fig:flat_sky}
\end{figure}
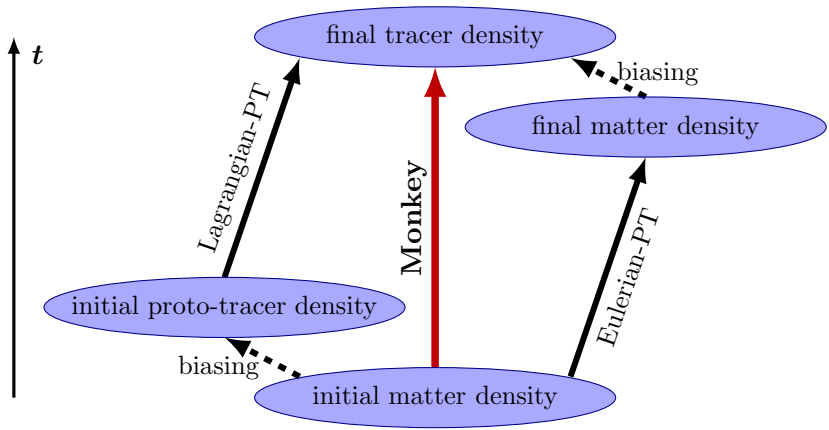
Canonically, this connection can be achieved either in the so called Eulerian or Lagrangian perturbative scheme.  
In Eulerian perturbative schemes 
\cite{McDonald:2006, McDonald+:2009, Chan++:2012, Baldauf++:2012, Saito++:2014, Assassi++:2014, Senatore:2014, Mirbabayi++:2014}
dark matter dynamics is treated as an effective fluid on the large scales 
governed by the continuity and the Euler equations with small scales providing backreaction in terms of the 
unknown source terms that can be organised perturbatively on large scales. Connection of dark matter to 
halos is, in this picture, achieved by writing, order by order in dark matter fields, the most general 
functional form allowed by the covariance under coordinate transformations. Thus, the Eulerian perturbative 
description of galaxy clustering, valid on quasi-linear scales, is based on two step procedure: 
\begin{itemize}
\item bias expansion of the tracer field in terms of the non-linear matter operators (density field, shear etc.).
\item Eulerian perturbation theory expansion of these non-linear matter operators.
\end{itemize}
This, two step procedure, is schematically shown on the right hand side of the Figure. \ref{fig:flat_sky}.%
\footnote{Mathematically, the $n$-th order biased tracer $\delta_h^{(n)}$ is first expanded 
by in operator basis of the non-linear matter field $\delta_m^n$ as $\delta_h^{(n)}=\sum_i b_i^{(n)}O_i^{(n)}$ 
where each operator in the sum is of order $\delta_m^n$, i.e. $O_i^{(n)}=\mathcal{O}(\delta_m^n)$. Once this step is done, each operator needs to be further expanded with respect to the linear density field 
$\delta_\ell^n$, as $O_i^{(n)}=\sum_jc_{ij}^{(n)} \tilde O_j^{(n)}$ where $\tilde O_j^{(n)}=\mathcal{O}(\delta_{\ell}^n)$.
The last step is nothing but the perturbation theory computation of the given nonlinear operator $O_i^{(n)}$.}
The Eulerian based biasing schemes is currently most often used for the 
description of bias tracer statistics, and even though various different set ups are employed \cite{McDonald+:2009,Senatore:2014, Mirbabayi++:2014} 
these are all mathematically equivalent, differing only on the level of the choice of the basis of operators (i.e. isomorphic). 
However, in all of these frameworks, only low order expansions in the EdS universe have been investigated 
(up to the third order in the biased field), and extension to the higher orders (see however \cite{Eggemeier:2018} for the one-loop bispectrum investigation)
or to the more generic background universe still remains somewhat undetermined. The construction of a 
basis at the $n-$th perturbative order is not streamlined and relies somewhat on the ingenuity and agility of the
investigator, without apparent means to ensure that the obtained bias operator basis is complete.
This is not to say that the higher-order biasing basis can not be constructed, but rather that these approaches do not allow for a streamlined, 
algorithmic and programable procedure, and thus have to resort to the `by hand' constructions.
Monkey bias basis, among other things, is constructed in particular to remedy these worries and to streamline 
the biasing basis construction order by order in the generic background universe, 
based solely on the physical principles and resulting constraints.

The Lagrangian schemes \cite{Matsubara:2008, Matsubara:2011, Carlson:2012, Matsubara:2013, 
Vlah++:2015, Vlah:2016, Aviles:2018, Vlah+:2018, Chen++:2020_I, Chen++:2020_II}
revert the two steps above. Here one starts by writing the biasing 
expansion for the so-called proto-halo field in some initial time-slice based on the same 
covariance principle. The second step relies on using the Lagrangian perturbation theory of dark matter 
displacements to advect this field to the final, Eulerian position.  The Lagrangian perturbative description of galaxy 
clustering, again constructed to be valid on quasi-linear scales, thus reverts these previous (Eulerian) steps: 
\begin{itemize}
\item bias expansion of the (proto-)tracer field is done in terms of the initial (linear) density distribution that is set up for evolution to the final time.
\item Lagrangian perturbation theory expansion of displacement field is used to evolve the biased tracer field.
\end{itemize}

These two schemes are shown in Figure.\ref{fig:flat_sky} and can represent all of the currently 
existing approaches to describing the connection of dark matter dynamics and the evolution of biased tracers. 
Crucially, the general bias expansion in either frame is mathematically equivalent, order by order in perturbation theory, 
if we assume the same physical conditions. The key feature of both of these schemes is the two-step procedure, 
where the nonlinear dynamics and the description of biased tracer fields undergo similar but seemingly unrelated perturbative expansion. 
One of the goals of the next section is to unify this into a single-step procedure, thus allowing us to optimize our description and computation. 

In general, the idea of the description of biased tracers statistics on the large scales relies on the assumption of scale separation.  
On large enough scales, larger than $R^*$, gravity is dominant and responsible for driving the dynamics of biased tracers. 
Below $R^*$, a variety of small scale physics, characteristic to a given tracer, is also active, while due to the scale separation, 
its impact on the large scales is suppressed. Galaxies can be considered as such tracers given the separation of scale criteria from above, 
despite the complex nature of galaxy formation and its nonlinear dynamics. 

The continuity equation for biased tracer is often assumed, which, of course, implies the number density of the biased tracer should 
be conserved.\footnote{In canonical constructions of bias basis, this equation is used to motivate the construction of the biasing 
operators along the lines of the standard perturbation theory (SPT). Once this procedure is done, one can relax the number density 
conservation requirement, 
by implicitly stating that an arbitrary source term 
in the continuity equation can also be expanded, order by order, in the same operator bases.}
In reality, this assumption is not entirely accurate for realistic bias tracers, since galaxies can undergo complex formation and 
merger dynamics that change their total number. Formally, we can represent these processes by adding source terms to the continuity equation. 
Similar considerations hold for the momentum balance, i.e., Euler equation, where, e.g., AGN outflows and feedbacks can affect the momentum balance of tracers. 
Therefore, we can write 
\begin{align}
\label{eq:EoM_tracers}
\partial_\tau \delta_h (\bm x, \tau) + {\bm \nabla} \cdot \left( \left[ 1 + \delta_h \right] \bm u_h \right)(\bm x, \tau) &= S_\delta (\bm x, \tau), \\
\partial_\tau \bm u_h (\bm x, \tau) + \mathcal H(\tau) \bm u_h (\bm x, \tau) + \bm u_h (\bm x, \tau) \cdot {\bm \nabla} \bm u_h (\bm x, \tau)   &= - {\bm \nabla} \phi (\bm x, \tau) + S_u (\bm x, \tau), \nonumber
\end{align} 
where the source terms $S_{\delta}$, $S_{u}$ encapsulate the small scale physics feedback on the 
large scale tracer density and velocity. The gravitational potential $\phi$ in the Euler equation is given 
by the Poisson equation in the Newtonian limit. Given that the gravitational interaction is dominant on large scales, 
a suitable model assumes that only density fields of cold dark matter and baryons contribute to the Poisson equation:
\begin{align}
\nabla^2 \phi (\bm x, \tau) &= \frac{3}{2} \mathcal H (\tau)^2 \Big( \Omega_c (\tau) \delta_c (\bm x, \tau) + \Omega_b (\tau) \delta_b (\bm x, \tau) \Big) \\
&=  \frac{3}{2} \mathcal H (\tau)^2  \Omega_m (\tau) \delta_m (\bm x, \tau),  \nonumber
\end{align} 
where $\bm x$ is the comoving position, $\tau$ is the conformal time
and $\Omega_c$ and $\Omega_b$ are the energy fraction of cold dark matter and baryons, respectively. 
We also have introduced the total matter fraction $\Omega_m = \Omega_c+\Omega_b$ and overdensity $\delta_m = (\Omega_c \delta_c + \Omega_b \delta_b)/\Omega_m$,
while the isocurvature perturbation is neglected.
The set of coupled equations is closed by  continuity and Euler equation for the dark matter fluctuation $\delta_m$ which are
the usual EFT equations for the long modes sourced by the small scale physics~\cite{Baumann++:2010, Carrasco++:2012, Carroll+:2013}. 

The source terms for biased tracers, $S_{\delta}$, $S_{u}$, can depend on their own field $\delta_h$ and velocity $\bm{u}_h$, 
the long fields of dark matter $\delta_m$ that drive the dynamics on large scales as well as on the small scale physics that is 
characterized by some typical $R^*$ scale.
Moreover, the processes can occur non-locally in time \cite{Senatore:2014, Mirbabayi++:2014, Desjacques:2016}, 
given that the galactic processes and dynamics occur on roughly Hubble time scales.
Thus, we can formally write the source terms as
\begin{align}
S_{\delta, u} (\bm x, \tau) = \int^\tau d\tau'~ s_{\delta, u} \left[ \delta_h, \bm u_h, \delta_m, \ldots, R^* \right] \left( \bm x_{\rm fl} , \tau' \right),
\label{source terms S}
\end{align} 
where $\bm x_{\rm fl}$ is the flow coordinate that can be expressed in terms of the Eulerian coordinate 
as $\bm x_{\rm fl}\left( \bm x, \tau, \tau' \right) = \bm x - \int_{\tau'}^\tau d\tau''~\bm u_h \left( \bm x_{\rm fl}\left( \bm x, \tau, \tau'' \right), \tau'' \right)$. 
Note that at this level of description we need not invoke the conventional notions that galaxies reside in massive dark matter dominated halos. 
In cases when this holds, this fact is reflected in the values of the bias coefficients indicating the level of correlation of dense dark matter 
region and the tracers at hand.

%====================================================================================%
\section{Monkey approach to description of biased tracers}
\label{sec:monkey}
%====================================================================================%

In this section, we present a novel approach to constructing the perturbative operator basis for biased tracers that we call the \textit{Monkey theory}.\footnote{
The name Monkey bias theory coagulated out of the numerous discussions amongst authors during witch various renderings of 
the \textit{infinite monkey theorem} \cite{wiki_IMT:2020} were invoked. Moreover, the algorithmic structure of the Monkey bias 
theory resembles the \textit{monkey software testing} technique \cite{wiki_MST:2020} used for user tests in software applications.
}
Instead of following the existing approaches of first building the list of allowed (and nonlinear) bias operators,
which subsequently have to be further expanded in PT, we first construct the superset bases at a given PT
order, which is then trimmed down to the final basis. The latter step is achieved by imposing the equivalence 
principle via consistency relations. 

%====================================================================================%
\subsection{General idea}
%====================================================================================%

Our goal is to find  a general expression for the density contrast of a biased tracer $\delta_h$ in terms of the linear density contrast of dark matter $\del$. 
In other words, we are interested in constructing the following perturbative expansion of the functional $\mathcal{F}$:
\begin{equation}
\delta_h(\bm{x},\tau) = \mathcal{F}[\del(\bm{x},\tau)].
\label{functional}
\end{equation}
Here, we adopt $\del (\bm{x})$ as the most fundamental building block, simply because it is the unique perturbative quantity whose property is robustly known. 
The above functional may be decomposed into the linear part and non-linear part,
\begin{equation}
\delta_h = a_1 \del + \mathcal{F}_{\rm NL}[\del],
\label{NL separation}
\end{equation}
where $a_1$ is the coefficient of the linear operator $\del$, and $\mathcal{F}_{\rm NL}$ includes all the non-linear operators.

The monkey theory consists of the following three steps in constructing the bias expansion $\mathcal{F}_{\rm NL}[\del]$ up to given order:
\begin{description}
\item[(Step 1)] 
Specifying all non-linear terms relevant to the evolution of the biased tracer.
\item[(Step 2)] 
Writing down all possible operators generated by these non-linear terms.
\item[(Step 3)]
Relating the coefficients of these operators by imposing physical requirements.
\end{description}
In the following subsections, we will explain each step and derive
the bias expansion up to the third order.
We drop the higher derivative and stochastic terms, represented in the source terms of eq.~\eqref{eq:EoM_tracers},
however the procedure we outline below can be extended in order to include also these operators, as discussed 
in Section~\ref{Inclusion of the higher derivatives and stochastic contributions}.

%====================================================================================%
\subsection{Step 1:  Specifying non-linear terms}
\label{step_1}
%====================================================================================%

The monkey theory requires non-linear terms in  the dynamical equation for the biased tracer as input. 
In other words, the monkey theory is a formalism which gives a general bias expansion for a given set 
of non-linear terms for the biased tracer.
Here, as a working example, we assume that the evolution equation for the biased tracer has the non-linear terms of the continuity equation and Euler equation (as in eq.~\eqref{eq:EoM_tracers}).
\begin{align}
{\rm Continuity\ eq.:}\quad\partial_\tau \delta+ {\rm (linear\ terms)} &=-\delta \theta-\partial_i\delta \frac{\partial_i}{\partial^2}\theta,
\label{Conteq}
\\
{\rm Euler\ eq.:}\quad
\partial_\tau\theta +{\rm (linear\ terms)} &=
-\frac{\partial_i\partial_j}{\partial^2}\theta
\frac{\partial_i\partial_j}{\partial^2}\theta,
\label{Eulereq}
\end{align}
where $\theta$ is the divergence of the velocity field, $\theta\equiv\bm{\nabla}\cdot \bm{v}$ and its curl component is ignored, $v_i =(\partial_i/\partial^2)\theta$. In the above equations, ``(linear terms)'' represents some linear terms of $\theta$ or $\delta$ which are irrelevant in the monkey theory.
Note that we do not necessarily assume that our biased tracer,
$\delta_h$ and $\theta_h$, satisfies the above equation. Instead, 
our assumption is that the nonlinear dynamics of the biased tracers 
is sourced by three non-linear terms, 
\begin{equation}
\big({\rm evolution\ eqs.\ for}\ \delta_h\big)\ \supset\
\Delta_h^2,
\ \partial_i \Delta_h \frac{\partial_i}{\partial^2}\Delta_h,
\ \frac{\partial_i\partial_j}{\partial^2}\Delta_h
\frac{\partial_i\partial_j}{\partial^2}\Delta_h,
\label{dh evolution}
\end{equation}
where $\Delta_h$ is either $\delta_h$ or $\theta_h$.\footnote{Note that,
motivated by the structure of nonlinear terms in Eq.~\eqref{Conteq} and Eq.~\eqref{Eulereq}.,
we consider here only quadratic couplings. One might wonder if extending the couplings with 
cubic and higher-order terms would generate new operators in the biasing basis. We explicitly 
checked that, up to the third order in perturbation theory used here, this is indeed not the case, 
i.e. adding such couplings produces only degenerate operators.   
}
Here we ignore the difference between $\delta_h$ and $\theta_h$,
because it will not change the resultant bias expansion and we would like to keep the non-linear terms as general as possible.
We do not try to determine the coefficients of these non-linear terms, because the coefficients of operators will be left as free 
parameters at the end in the bias expansion approach. 
We note that the second generator $\partial_i \Delta_h \frac{\partial_i}{\partial^2}\Delta_h$ is related to displacements,
and as such it should be constructed from the matter density field. This is obviously a consequence of the equivalence principle
and the fact that at the large-scale displacement field of matter and biased tracer should be the same.
However we do not need to impose these requirements since this is precisely the content of the consistency relations, 
and thus these constraints will naturally follow by imposing them in Step 3.

Now we take a further step towards abstraction and focus only 
on the mathematical structure of  eq.~\eqref{dh evolution}, %
\begin{equation}
\left\{ X Y,\quad 
\partial_i X\frac{\partial_i}{\partial^2}Y,\quad 
\frac{\partial_i\partial_j}{\partial^2} X\frac{\partial_i\partial_j}{\partial^2} Y\right\},
\label{NL terms}
\end{equation}
where $X$ and $Y$ are arbitrary operators.
The point of eq.~\eqref{NL terms} is that when we perturbatively solve the biased tracer evolution and obtain the $(n+1)$-th
order solution, such a solution must be generated from  the $n$-th order solution which are substituted into the non-linear source term in eq.~\eqref{dh evolution}.
For instance, since we know $\delta_h\propto \del$ at the first order (see eq.~\eqref{NL separation}), 
the second order solution which is generated through the non-linear terms in eq.~\eqref{dh evolution}
should be a linear combination of the following three operators,
\begin{equation}
\del \del,\quad 
\partial_i \del\frac{\partial_i}{\partial^2}\del,\quad 
\frac{\partial_i\partial_j}{\partial^2} \del\frac{\partial_i\partial_j}{\partial^2} \del.
\label{2nd operator}
\end{equation}
We can continue to construct the bias basis at higher orders in this way as we see below.
Since $\del(\bm x,\tau)$ has its time and spatial dependence in a separable form, these operators satisfy the same property 
and we can focus on their spatial dependence, because the time dependence will be inevitably degenerated with that of the bias parameters.

Before finishing the step 1, let us stress again that 
the list of the non-linear terms in eq.~\eqref{dh evolution}
is an input for the monkey theory. Even if the list changes, 
our formalism still works and provides another bias
expansion. 
Since the matter fluctuation $\delta_m$ itself can be regarded as a biased tracer (be it a trivial one), we assumed dark matter also satisfies requirement 
in ~\eqref{dh evolution} but it is the only assumption on the matter dynamics so far. However, unlike the other formalisms assuming a concrete matter evolution 
(e.g. SPT), our theory is agnostic on the details of its evolution as long as it conforms to the structure posed in eq.~\eqref{dh evolution}.

%====================================================================================%
\subsection{Step 2: Writing down all possible operators}
\label{step_2}
%====================================================================================%

Provided that we only have the three non-linear terms with structures of eq.~\eqref{NL terms} in the dynamics of biased tracer, 
any non-linear operators which appear in the bias functional $\mathcal{F}_{\rm NL}[\del]$ should be generated through eq.~\eqref{NL terms}.
Therefore, we find the following generating rule of operators:
\begin{align}
O^{(n)}&=  
\left\{O^{(m)}O^{(n-m)},\ 
\partial_i O^{(m)}\frac{\partial_i}{\partial^2}O^{(n-m)},\ 
\frac{\partial_i\partial_j}{\partial^2} O^{(m)}\frac{\partial_i\partial_j}{\partial^2} O^{(n-m)}
\right\},
~ m=1,\dots,n-1
\\
O^{(1)}&=\delta_\ell,
\end{align}
where $O^{(n)}$ denotes an operator at $n$-th order of the perturbation.
One should exhaust all possible operators at $n$-th order by 
substituting all the combinations of $O^{(m)}$ and $O^{(n-m)}$
for $m=1,2,..., n-1$. From this rule, we easily obtain all the second and third order operators.
Assigning independent free coefficients to them, 
the tentative bias expansion up to the third order is written as%
\footnote{Here it is understood that the second order terms in bias expansion are subtracted by their expectation values (e.g. $\del^2\to  \del^2-\langle\del^2\rangle$)
in order to ensure $\langle \delta_h\rangle=0$. 
Ignoring primordial non-gaussianity, we assume that $\del$ is a gaussian random field.}
\begingroup
\allowdisplaybreaks
\begin{align}
\label{deltah list}
\delta_h&= a_1 \del
%\quad +({\rm higher\ derivative\ terms})+({\rm stochastic\  terms})
\\
&+b_1 \del^2+b_2 \partial_i\del\frac{\partial_i}{\partial^2}\del
+b_3 \frac{\partial_i \partial_j}{\partial^2}\del\frac{\partial_i \partial_j}{\partial^2}\del
\notag\\
&+c_1 \del^3 +c_2 \del\partial_i\del\frac{\partial_i}{\partial^2}\del
+c_3 \del\frac{\partial_i \partial_j}{\partial^2}\del\frac{\partial_i \partial_j}{\partial^2}\del
\notag\\
&+d_1 \partial_i\del\frac{\partial_i}{\partial^2}\left(\del^2 \right)
+d_2 \partial_i\del\frac{\partial_i}{\partial^2}\left(\partial_j\del\frac{\partial_j}{\partial^2}\del \right)+d_3 \partial_i\del\frac{\partial_i}{\partial^2}\left(\frac{\partial_j \partial_k}{\partial^2}\del\frac{\partial_j \partial_k}{\partial^2}\del \right)
\notag\\
&+e_1 \frac{\partial_i}{\partial^2}\del\partial_i\left(\del^2 \right)
+e_2 \frac{\partial_i}{\partial^2}\del\partial_i\left(\partial_j\del\frac{\partial_j}{\partial^2}\del \right)
+e_3 \frac{\partial_i}{\partial^2}\del\partial_i\left(\frac{\partial_j \partial_k}{\partial^2}\del\frac{\partial_j \partial_k}{\partial^2}\del \right)
\notag\\
&+f_1 \frac{\partial_i\partial_j}{\partial^2}\del\frac{\partial_i\partial_j}{\partial^2}\left(\del^2 \right)
+f_2 \frac{\partial_i\partial_j}{\partial^2}\del\frac{\partial_i\partial_j}{\partial^2}\left(\partial_k\del\frac{\partial_k}{\partial^2}\del \right)
+f_3 \frac{\partial_i\partial_j}{\partial^2}\del\frac{\partial_i\partial_j}{\partial^2}\left(\frac{\partial_k \partial_l}{\partial^2}\del\frac{\partial_k \partial_l}{\partial^2}\del \right), \notag
\end{align}
\endgroup
where $a_1, b_1, b_2 \dots, f_3$ are coefficients and $\del$ scales with the linear growth rate $D(\tau)$ (proportional to the scaling factor $a(\tau)$ in the EdS).
Here, we suppressed the higher derivative terms and the stochastic terms, while they will be discussed in Section.~\ref{Inclusion of the higher derivatives and stochastic contributions}.
Note that the $c_2$ term and $e_1$ term are equivalent and we eliminate the $e_1$ term henceforth.
At this point we have 15 operators, and thus apparently have as many as 15 free parameters.
However, as we see in the next subsection, by requiring physical conditions one finds that 8 parameters out of these 15 are not free.

From its construction it is obvious that any perturbative formalisms 
whose non-linear terms fall into the expression of eq.~\eqref{dh evolution} give solutions in the above form.
For instance, the SPT solution in EdS universe for the dark matter density contrast $\delta_m$ is given by
\begin{align}
\label{SPT EdS}
&a_1^{(m)}=1,\, b_1^{(m)}=\frac{5}{7},\, b_2^{(m)}=1,\, b_3^{(m)}=\frac{2}{7},\,
c_1^{(m)}=\frac{4}{9},\, c_2^{(m)}=\frac{10}{7},\, c_3^{(m)}=\frac{1}{3},\,
\\ 
&d_1^{(m)}=\frac{3}{14},\, d_2^{(m)}=\frac{1}{2},\, d_3^{(m)}=\frac{2}{7}, e_2^{(m)}=\frac{1}{2},\, e_3^{(m)}=\frac{11}{63},\,
f_1^{(m)}=\frac{2}{21},\, f_2^{(m)}=\frac{2}{9},\, f_3^{(m)}=\frac{8}{63}. \notag
\end{align}
The case of the $\Lambda$CDM universe can be also found in appendix~\ref{app:lcdm_bias}.
Note that $a_1^{(m)}$ is always unity by definition.

%====================================================================================%
\subsection{Step 3: Relating the coefficients}
\label{Step 3: Relate the coefficients}
%====================================================================================%

So far, we did not solve any physical evolution but used only the mathematical structure of the non-linear terms in eq.~\eqref{dh evolution}. 
Thus, our bias expansion, given in eq.~\eqref{deltah list}, is too general, and it includes unphysical solutions. In order to constrain the parameter 
space into the physical one, we now impose conditions, namely the consistency relations of large scale structure, which are derived 
from the equivalence principle and adiabatic perturbation 
\cite{Kehagias+:2013, Peloso+:2013a, Creminelli++:2013a, Peloso+:2013b, Creminelli++:2013b, Horn:2014}. 
Since, in this paper, we are interested in the bias expansion up to the third order, we consider 
3-point and 4-point correlators at tree-level (as well as power spectrum at the one-loop level in Appendix \ref{app:loop_ps}).

%====================================================================================%
\subsubsection{3-point function}
%====================================================================================%

According to the consistency relations,  the unequal-time correlator between 
$n$ biased tracers $\delta_{h_i}(\bm{k}_i,\tau_i)\ (i=1,\cdots,n)$ and one matter fluctuation $\delta_m(\bm q, \tau)$ with a soft momentum $q\ll k_i$
 has a leading contribution of $\mathcal{O}(q^{-1})$, 
\begin{align}
\label{general single limit}
\lim_{q\to 0}
&\langle\delta_m(\bm q,\tau) \delta_{h_1}(\bm k_1,\tau_1) 
\delta_{h_2}(\bm k_2,\tau_2)\cdots \delta_{h_n}(\bm k_n,\tau_n) \rangle'
\\
&\hspace{2cm}\approx -P_m(q,\tau)\left(\sum_{i=1}^n \frac{D(\tau_i)}{D(\tau)}\frac{\bm k_i\cdot \bm q}{q^2}\right)
\langle \delta_{h_1}(\bm k_1,\tau_1) 
\delta_{h_2}(\bm k_2,\tau_2)\cdots \delta_{h_n}(\bm k_n,\tau_n) \rangle', \notag
\end{align}
where $h_i$ can be all different species of biased tracers,  
%$D(\tau)$ is the growth factor of density fluctuation 
and $P_m$ is the power spectrum of the long mode $\delta_m(\bm q,\tau)$.
In the equal time limit, $\tau_1=\tau_2=\cdots=\tau_n$, the summation over the momenta $\bm k_i$ cancels out due to the momentum conservation and 
the leading contribution of $\mathcal{O}(q^{-1})$ vanishes.
Therefore, the consistency relation requires the equal-time correlator
to be free from IR-divergence
%\footnote{Strictly speaking, logarithmic divergences are not prohibited %in this argument. However, the exclusion of $\mathcal{O}(q^{-1})$ is sufficient %for our purpose.}
%
\begin{equation}
\lim_{q\to 0}
\frac{\langle\delta_m(\bm q,\tau) \delta_{h_1}(\bm k_1,\tau) 
\delta_{h_2}(\bm k_2,\tau)\cdots \delta_{h_n}(\bm k_n,\tau) \rangle'}
{P_m(q,\tau)\langle \delta_{h_1}(\bm k_1,\tau_1) 
\delta_{h_2}(\bm k_2,\tau_2)\cdots \delta_{h_n}(\bm k_n,\tau_n) \rangle'}\not\owns \mathcal{O}(q^{-1}).
\end{equation}
Henceforth, we shall suppress time arguments when we consider equal-time correlators.

Let us apply this requirement to our bias expansion.
Tree-level 3-point correlator between two different bias tracers $h_\alpha$ and $h_\beta$ and matter fluctuation in the soft limit is computed as 
\begin{equation}
\lim_{q\to0}\langle\delta_m(\bm q)\delta_{h_\alpha}(\bm k_1) \delta_{h_\beta}(\bm k_2) \rangle_{\rm tree}'
=\left(a_1^{(\alpha)} b_2^{(\beta)}-a_1^{(\beta)}b_2^{(\alpha)}\right)
\frac{\bm q \cdot \bm k_1}{2q^2} P_\ell(q) P_\ell(k_1)+\mathcal{O}(q^0).
\end{equation}
The first term in the right hand side is IR-divergent $\mathcal{O}(q^{-1})$.
Hence the consistency relations generally impose 
$a_1^{(\alpha)} b_2^{(\beta)}-a_1^{(\beta)}b_2^{(\alpha)}=0$
for arbitrary biased tracers $\alpha$ and $\beta$. 
This condition implies that any biased tracer satisfies the following equation with an universal coefficient $\mathcal{C}_{b}$,
\begin{equation}
\frac{b_2^{(\alpha)}}{a_1^{(\alpha)}} = \frac{b_2^{(\beta)}}{a_1^{(\beta)}}=
\mathcal{C}_{b}.
\label{b2 condition}
\end{equation}
We label the coefficient $\mathcal{C}_{b}$ is ``universal'' in the sense of being independent of the tracer species 
and in general just a function of time.  It can be fixed by imposing concrete equations of motion for some tracer, specifically, the usual SPT solution for dark matter. 
Indeed, assuming the SPT solution for dark matter in the EdS universe,
one finds $\mathcal{C}_{b}=1$.

The above constraint thus imply that the coefficient of the operator $\partial_i\del\frac{\partial_i}{\partial^2}\del$ is 
not free, but linked to the coefficient of the linear term $\del$ for any tracer. This is, of course, a known result saying that 
the displacement of the fields does not change the linear bias parameter. %that multiplied the linear field. 
This constraint is, by construction, also present in the other formalisms which we will discuss in Section.~\ref{sec:comp}~\cite{McDonald+:2009,Senatore:2014, Mirbabayi++:2014}. 
In our case, it follows purely as a requirement of the  consistency relation. Up to the second order in the tracer field we thus end 
up with just two independent operators $\del^2$ and $\frac{\partial_i \partial_j}{\partial^2}\del\frac{\partial_i \partial_j}{\partial^2}\del$
with free bias coefficients, $b_1$ and $b_3$ respectively. 
%

%====================================================================================%
\subsubsection{4-point function}
%====================================================================================%

In a similar way, tree-level 4-point correlator between three different bias tracers, $\alpha,\beta,\gamma$, and matter fluctuation in the soft limit,
$\lim_{q\to 0}\langle\delta_{m}(\bm q)\delta_{h_{\alpha}}(\bm k_1) \delta_{h_{\beta}}(\bm k_2) \delta_{h_{\gamma}}(\bm k_3)\rangle_{\rm tree}'$,
should not include IR-divergent terms $\mathcal{O}(q^{-1})P_\ell(q)
\langle\delta_{h_{\alpha}}(\bm k_1) \delta_{h_{\beta}}(\bm k_2) \delta_{h_{\gamma}}(\bm k_3)\rangle_{\rm tree}'$.
One can show that the required conditions are
\begin{equation}
c_2^{(h)}=2\mathcal{C}_{b}b_1^{(h)},\qquad 
d_2^{(h)}=e_2^{(h)}=\frac{1}{2}\mathcal{C}_{b}^2 a_1^{(h)},\qquad 
f_2^{(h)}+2e_3^{(h)}=2\mathcal{C}_{b} b_3^{(h)},
\label{single IR}
\end{equation}
where $\mathcal{C}_{b}$ is the universal coefficient introduced in eq.~\eqref{b2 condition}.
This gives us four new constraints, fixing four third order bias coefficients. 
In addition, 4-point correlator allows us to take other two kinds of soft limits. One is double soft limit of external momenta.
4-point correlator with two soft legs should satisfy
\begin{align}
\lim_{q_1,q_2\to 0}\frac{\langle\delta_{m}(\bm q_1) \delta_{m}(\bm q_2)\delta_{h_{\alpha}}(\bm k_1) \delta_{h_{\beta}}(\bm k_2)\rangle'}{P_m(q_1)P_m(q_2)
\langle\delta_{h_{\alpha}}(\bm k_1) \delta_{h_{\beta}}(\bm k_2)\rangle'}
\not\owns \mathcal{O}(q^{-1}_1,q_2^{-1}).
\end{align}
It can be shown that this correlator at tree-level with our bias expansion yields terms
with $\mathcal{O}(q^{-2})$ and $\mathcal{O}(q^{-1})$ where $q\sim q_1, q_2$.
The condition to vanish the $\mathcal{O}(q^{-2})$ terms is 
\begin{equation}
b_{2}^{(\alpha)}b_{2}^{(\beta)} - a_1^{(\alpha)} e_2^{(\beta)}- a_1^{(\beta)} e_2^{(\alpha)}=0,
\label{soft two legs}
\end{equation}
which always holds with the second condition in eq.~\eqref{single IR}, and thus it does not provide us with the 
additional constraints on the third order bias coefficients.  
On the other hand, the condition to erase the $\mathcal{O}(q^{-1})$ terms provides new relations, 
$d_1^{(\alpha)}/a_1^{(\alpha)}=d_1^{(\beta)}/a_1^{(\beta)}$ and $d_3^{(\alpha)}/a_1^{(\alpha)}=d_3^{(\beta)}/a_1^{(\beta)}$.
Therefore, we introduce two, third order, universal coefficient
\begin{equation}
\frac{d_1^{(h)}}{a_1^{(h)}}=\mathcal{C}_{d},
\qquad
\frac{d_3^{(h)}}{a_1^{(h)}}=\tilde{\mathcal{C}}_{d}.
\label{double IR}
\end{equation}
As in case of the second order universal coefficient $\mathcal{C}_{b}$, 
the third order universal coefficients $\mathcal{C}_{d}$ and $\tilde{\mathcal{C}}_{d}$, are independent of the bias species and 
they again link the third order displacements to the linear bias coefficient of the linear operator.
As we did for the second order coefficient, we can again specify the matter dynamics by the SPT solutions which provides values of the 
$\mathcal{C}_{d}$ and $\tilde{\mathcal{C}}_{d}$. Assuming the EdS universe we get $\mathcal{C}_{d}=3/14$ and 
$\tilde{\mathcal{C}}_{d} = 2/7$. 

The last soft limit of 4-point correlator at tree-level is so-called collapsed limit in which the sum of two external momenta becomes very small.
Its consistency relation is written as
\begin{align}
\lim_{q_1,q_2\to 0}\frac{\langle\delta_{h_{\alpha}}(\bm k_1) \delta_{h_{\beta}}(\bm q_1-\bm k_1)\delta_{h_{\gamma}}(\bm k_2) 
\delta_{h_{\delta}}(\bm q_2-\bm k_2)\rangle'}{P_m(q_1)\langle\delta_{h_{\alpha}}(\bm k_1) \delta_{h_{\beta}}(-\bm k_1)
\rangle\langle \delta_{h_{\gamma}}(\bm k_2) \delta_{h_{\delta}}(\bm -\bm k_2)\rangle'}
\not\owns \mathcal{O}(q^{-1}_1,q_2^{-1}).
\end{align}
This time, both  $\mathcal{O}(q^{-2})$ and $O(q^{-1})$ terms automatically vanish under the conditions obtained so far.  
Nevertheless, we find that the dependence on the soft momentum $\bm q$
remains at $\mathcal{O}(q^0)$ order,
\begin{align}
\label{trispectrum_new_terms}
&\lim_{q\to 0}\langle\delta_{h_{\alpha}}(\bm k_1) \delta_{h_{\beta}}(\bm q-\bm k_1)\delta_{h_{\gamma}}(\bm k_2) \delta_{h_{\delta}}(-\bm q-\bm k_2)\rangle'
\\
&=2 a_1^{(\gamma)}a_1^{(\delta)}
\left[a_1^{(\beta)}\left(f_1^{(\alpha)}-f_2^{(\alpha)}+f_3^{(\alpha)}\right)
+a_1^{(\alpha)}\left(f_1^{(\beta)}-f_2^{(\beta)}+f_3^{(\beta)}\right)\right]
P_\ell(k_1)P_\ell^2(k_2)(\hat{\bm k}_1\cdot \hat{\bm q})^2
\notag\\
&~~+2 a_1^{(\alpha)}a_1^{(\beta)}
\left[a_1^{(\delta)}\left(f_1^{(\gamma)}-f_2^{(\gamma)}+f_3^{(\gamma)}\right)
+a_1^{(\beta)}\left(f_1^{(\delta)}-f_2^{(\delta)}+f_3^{(\delta)}\right)\right]
P_\ell^2(k_1)P_\ell(k_2)(\hat{\bm k}_2\cdot \hat{\bm q})^2
\notag\\
&~~+ a_1^{(\alpha)}a_1^{(\beta)} a_1^{(\gamma)}a_1^{(\delta)}
\Big[\mathcal{C}_{b}^2-2\big(\mathcal{C}_{d} +\tilde{\mathcal{C}}_{d}\big)\Big]
\left(k_1P_\ell'(k_1)P_\ell^2(k_2)(\hat{\bm k}_1\cdot \hat{\bm q})^2 \right. \notag\\
&\hspace{9.5cm} \left. + k_2P_\ell^2(k_1)P_\ell'(k_2)(\hat{\bm k}_2\cdot \hat{\bm q})^2\right)+\cdots, \notag
\end{align}
where $P_\ell(k)$ and $P_\ell'(k)$ are the linear matter power spectrum and its derivative, and the suppressed terms in $\cdots$ do not depend on $\hat{\bm q}$.
Requiring this $\hat{\bm q}$ dependence to vanish leads to 
\begin{equation}
f_1^{(h)}-f_2^{(h)}+f_3^{(h)}=0,
\qquad
\mathcal{C}_{b}^2-2\big(\mathcal{C}_{d} +\tilde{\mathcal{C}}_{d}\big)=0.
\label{f123}
\end{equation}
As shown in Appendix~\ref{app:loop_ps}, the same conditions are derived by 1-loop power spectrum (see eq.~\eqref{1loop f123}).
First constraint above reduces the number of free bias parameters, locking the three bias operators
in the last line of eq.~\eqref{deltah list}. On the other hand, the second constraint above, gives us the 
relation amongst the second and third order universal coefficients $\mathcal{C}_{b}, \mathcal{C}_{d}$ and $\tilde{\mathcal{C}}_{d}$. 
Using this constraint we can thus eliminate one of the third order coefficients. Note that for their values in the case with the 
SPT solution in the EdS quoted above, this condition is automatically satisfied (also in the generalized $\Lambda$CDM case, see eq.~\eqref{SPT LCDM}).

It is important to stress that these two conditions do not follow directly from the current form of the consistency conditions 
\cite{Kehagias+:2013, Peloso+:2013a, Creminelli++:2013a, Peloso+:2013b, Creminelli++:2013b, Horn:2014}
and thus constitute, at this stage, a separate and independent condition. However, it is evident that such contributions 
are unphysical and thus are most likely related to some additional symmetry or physical principle. 
These should then also be imposed on our system in addition to the equivalence principle. 
One can also consider these from the other perspective. In e.g.~\cite{Takada+:2013, Akitsu++:2016, Barreira++:2017} 
(see also the references therein) one writes trispectrum contributions that are relevant for determining the
power spectrum covariance as
\begin{equation}
 \lim_{q\to 0} T \left( \bm k_1, -\bm k_1+\bm q, \bm k_2,- \bm k_2- \bm q \right) 
 = T_c^{NG} \left( \bm k_1, -\bm k_1, \bm k_2,- \bm k_2 \right) +
 T^{SSC} \left( \bm k_1, -\bm k_1, \bm k_2,- \bm k_2; \bm q \right),
\end{equation}
where $T_c^{NG}$ is the connected non-Gaussian part of the trispectrum depending only on 
the hard momenta $\bm k_1$ and $\bm k_2$. $T^{SSC}$ is so-called super-sample covariance arising due to large-scale correlations of the modes with soft momentum $\bm q$, and is proportional to $P_\ell(q)$. 
In our expression eq.~\eqref{trispectrum_new_terms}, we see that the presented terms do not conform to this structure and hence are required to vanish. However, we caution the reader that this requirement is basically equivalent to the 
one we made above. It still requires the link to some underlying physical principle, 
as is done with the previous consistency conditions. We shall return to this issue in more detail in the followup work.

Finally, we note that the constraints obtained from the two- and three-point functions are consistent with 
imposing the consistency relations on the one-loop power spectrum. This comparison is done in Appendix~\ref{app:loop_ps}.
This is to be expected, given that the momenta configurations entering the kernels at one-loop power spectrum are the 
subset of contributions considered when exploring the limits of two- and three-point functions. This reasoning can be extended 
to the higher-order functions and higher loops. Therefore, extracting the constraints from tree-level statistics order by order 
is sufficient to ensure the IR safety of all the relevant loop contributions.

%====================================================================================%
\subsection{Result}
%====================================================================================%

Collecting all the constraints we are able to reduce number of bias parameters 
by one at the second order and by seven at the third order. 
Putting it all together, we obtain our physical Monkey bias expansion,
\begingroup
\allowdisplaybreaks
\begin{align}
\label{monkey_bias}
\delta_h&= \Red{a_1}\Bigg[\del+\frac{\Blue{\mathcal{C}_{b}^2}}{2} \frac{\partial_i}{\partial^2}\del\partial_i\left(\partial_j\del\frac{\partial_j}{\partial^2}\del \right)
\\
&\qquad\quad+ \partial_i\del\frac{\partial_i}{\partial^2}
\Bigg\{\Blue{\mathcal{C}_{b}}\del+\Blue{\mathcal{C}_{d}} \del^2 
+\frac{\Blue{\mathcal{C}_{b}^2}}{2} \partial_j\del\frac{\partial_j}{\partial^2}\del
+\left(\frac{\Blue{\mathcal{C}_{b}^2}}{2}-\Blue{\mathcal{C}_{d}}\right)\frac{\partial_j \partial_k}{\partial^2}\del\frac{\partial_j \partial_k}{\partial^2}\del\Bigg\}\Bigg]
\notag \\
&+\Red{b_1} \del \left[1 +2\Blue{\mathcal{C}_{b}} \partial_i\del\frac{\partial_i}{\partial^2}\right]\del
+\Red{b_3} \left[1+\Blue{\mathcal{C}_{b}}\frac{\partial_k}{\partial^2}\del\partial_k \right] \left( \frac{\partial_i \partial_j}{\partial^2}\del\frac{\partial_i \partial_j}{\partial^2}\del\right)
\notag\\
&+\Red{c_1} \del^3 
+\Red{c_3} \del\frac{\partial_i \partial_j}{\partial^2}\del\frac{\partial_i \partial_j}{\partial^2}\del
+\Red{f_1} \frac{\partial_i\partial_j}{\partial^2}\del\frac{\partial_i\partial_j}{\partial^2}\left(\del^2 -\frac{\partial_k \partial_l}{\partial^2}\del\frac{\partial_k \partial_l}{\partial^2}\del\right)
\notag\\&
+\Red{f_2} \left[\frac{\partial_i\partial_j}{\partial^2}\del\frac{\partial_i\partial_j}{\partial^2}\left(\partial_k\del\frac{\partial_k}{\partial^2}\del +\frac{\partial_k \partial_l}{\partial^2}\del\frac{\partial_k \partial_l}{\partial^2}\del\right)
-\frac{1}{2} \ \frac{\partial_i}{\partial^2}\del\partial_i\left(\frac{\partial_j \partial_k}{\partial^2}\del\frac{\partial_j \partial_k}{\partial^2}\del \right)\right]
\notag
\end{align}
\endgroup
where we omit the superscript $(h)$ of the coefficients.
The bias parameters and the universal coefficients are highlighted by \Red{red} and \Blue{blue} color, respectively.
Consequently, the bias expansion of $\delta_h$ has one, two and four free parameters at first, second and third perturbative order, respectively,
\begin{equation}
{7\ \rm free\ parameters}:\quad 
{\rm [1st\ order]}\ \ \Red{a_1},\ \
{\rm [2nd\ order]}\ \ \Red{b_1}, \Red{b_3},\ \ 
{\rm [3rd\ order]}\ \ \Red{c_1}, \Red{c_3}, \Red{f_1}, \Red{f_2}.
\notag
\end{equation}
We remind the reader that these are only the `deterministic'
and the lowest-order derivative bias operators and that one should add to these appropriate 
stochastic and higher derivative operators in order to obtain the full expansion of the biased field. We return to these in the 
subsection below. 

Besides the bias parameters in our Monkey bias basis in eq.~\eqref{monkey_bias}, we also have $\Blue{\mathcal{C}_{b}}$ and $\Blue{\mathcal{C}_{d}}$, 
which were introduced as universal coefficients in eqs.~\eqref{b2 condition} and \eqref{double IR} (and after using eq.~\eqref{f123}), respectively.
If one assumes to know the non-linear dynamics of the matter fluctuation (or any other biased tracer), these coefficients can be fixed using the relations 
$\mathcal{C}_{b}=b_2^{(m)}/a_1^{(m)}$ and $ \mathcal{C}_{d}=d_1^{(m)}/a_1^{(m)}$ obtained by imposing the consistency relations. 
If matter dynamics can be described by the SPT in the EdS universe, for instance, $\mathcal{C}_{b}=1$ and $\mathcal{C}_{d}=3/14$ 
are true for all biased tracers.
Thus, $\mathcal{C}_{b}$ and $\mathcal{C}_{d}$ are not counted as bias parameters, given that they do not depend on the 
characteristics of a specific tracer. 
In the $\Lambda$CDM universe, the coefficients are no longer constant but have some fixed time-dependence.
Using the results given in Appendix~\ref{app:lcdm_bias}, we get $\mathcal{C}_{b}=1$ and $\mathcal{C}_{d}=(3 \lambda_1-12 \lambda_2+3 \nu_3-2)/8$,
where $\lambda_n$ and $\nu_n$ are time dependent functions encapsulating the influence of the cosmic expansion on 
the structure formation \cite{Fasiello+:2016a, Fasiello+:2016b}.

Measuring the $n$-point correlators within reach of the expansion given in eq.~\eqref{monkey_bias}  (up to the four-point function) allows us 
to measure, in principle, the $\mathcal{C}_{b}$ and $\mathcal{C}_{d}$ directly from the statistics of biased tracers. 
Moreover, as done for the other bias parameters, near-optimal estimators can also be constructed for these coefficients.
Measuring the deviations form the $\Lambda$CDM values could provide tests of GR and could indicate the presence 
of new physics (e.g., a new degree of freedom). For example, in simplified models where quintessence fluctuations 
are present \cite{Sefusatti+:2011,Anselmi++:2011}, these universal coefficients are sensitive to the deviation from 
the $\Lambda$CDM universe as $\mathcal{C}_{b}=1-\epsilon_1$ and $\mathcal{C}_{d}= (3 \lambda_1-12 \lambda_2+3 \nu_3-2+2\epsilon_2)/8$
where the $\epsilon_1$ and $\epsilon_2$ parameters represent the deviation
 (see Appendix~\ref{app:lcdm_bias} for all the coefficients and \cite{Fasiello+:2016a} for the definition of $\epsilon_n$). 
 Thus, they provide a potentially clean signal beyond GR, which is non-degenerate with the rest of the bias coefficients. 
Note that these particular models satisfy the consistency relations \cite{Fasiello+:2016b, Lewandowski+:2017, Fasiello+:2018, Lewandowski:2019}, 
even though the time evolution of the displacement operators is modified.
Constructing near-optimal and optimal estimators along the lines of 
\cite{Schmittfull++:2014, Lazeyras+:2017, Abidi+:2018, MoradinezhadDizgah++:2019}, could be 
a promising path in obtaining the constraints on the time evolution of the coefficients and its possible modifications. 

%====================================================================================%
\subsection{Inclusion of the higher derivatives and stochastic contributions}
%====================================================================================%
\label{Inclusion of the higher derivatives and stochastic contributions}

As noted above, so far, we have restricted our analysis in considering only deterministic and the lowest-order derivative 
terms of the bias operators. This restriction is not necessary and can be lifted relatively easily by extending the Monkey 
basis in eq.~\eqref{deltah list} to include appropriate stochastic and derivative operators. However, in order to include 
these additional operators, we need to choose the corresponding power counting.
It is usual practice to add the stochastic operators already at the linear level, as is appropriate, 
while the leading derivative operator, $R^2_*\partial^2\del$, suppressed by some scale $R_*$, starts to contribute at the third field order.

In eq.~\eqref{functional}, we have assumed that the biased tracer $\delta_h(\bm{x},\tau)$ depends only on the linear matter 
fluctuation $\del(\bm{x},\tau)$. In reality, however, the biased tracer $\delta_h(\bm{x},\tau)$ is not entirely determined by 
$\del(\bm{x},\tau)$ on large scales, given that it also depends on the small scale dynamical modes of dark matter as 
well as baryonic physics also present at small scales. Due to our limited knowledge of the small scale dynamics,
the biased tracers have an un-deterministic component on top of the biased expansion with respect to $\del(\bm{x},\tau)$ 
in eq.~\eqref{deltah list}. In order to make up this gap, one can introduce a new stochastic field $\epsilon(\bf x, \tau)$ which is 
uncorrelated with $\del(\bm{x},\tau)$. With this stochastic field, the starting point of our discussion in Sec.~\ref{sec:monkey}
is extended as
\begin{equation}
\label{stochastic}
\delta_h = \mathcal{F}[\del,\epsilon]=a_1 \del +
a_\epsilon \epsilon+
\mathcal{F}_{\rm NL}[\del,\epsilon],
\end{equation}
where $a_\epsilon$ is the linear bias coefficient of the stochastic term.
It is straightforward to start over from the Step 1 (subsec.~\ref{step_1}) 
to obtain the nonlinear bias expansion $\mathcal{F}_{\rm NL}[\del,\epsilon]$ by allowing the $X$ and $Y$ 
variables in eq.~\eqref{NL terms} to take values of the new stochastic field $\epsilon(\bf x, \tau)$.
The new stochastic field $\epsilon$ can be characterized, as usual, by having only constant Fourier space 
correlators and no cross-correlations with any of the deterministic operators.

In principle, higher-order operators containing combinations of deterministic and stochastic operators 
do not depend only on one and the same stochastic field $\epsilon$ but can each have a new degree of  
freedom in terms of the new stochastic fields $\epsilon_O$ (see, e.g. \cite{Mirbabayi++:2014, Angulo++:2015, Desjacques:2016}). 
This is, in principle, simple to incorporate in the algorithm above by attaching an additional label to the stochastic 
field at each iteration step, in effect proclaiming it a new, independent, stochastic field. In this way, the full freedom 
of stochastic contributions can be covered.
However, note that much of this is already captured in the current form given in eq.~\eqref{stochastic}.
Given that we have no a priory knowledge of the distribution nor dynamics of the stochastic field, for every correlator 
$\langle \epsilon \epsilon \rangle$, $\langle \epsilon \epsilon \epsilon \rangle$, etc., we require new functional freedom. 
This alone results in functional freedom in the structure of correlators of the stochastic field that would be quite degenerate 
with the form obtain if multiple stochastic fields would be used, as suggested above.

The higher derivative terms can be  added by extending the equation
of motion for the biased tracer, eq.~\eqref{dh evolution}.
As discussed around eq.~\eqref{eq:EoM_tracers}, 
the equations of motion for $\delta_h$ include the source terms
eq.~\eqref{source terms S} brought up by the small scale physics feedback on the large scale tracer.
Using the Taylor expansion, one finds that the leading order contribution
from the source terms is approximately given by
\begin{equation}
S_{\delta, u} (\bm x, \tau) \sim R_*^2 \partial^2 \del(\bm x, \tau).
\end{equation}
Since this derivative term is suppressed by the small scale parameter $R_*$,
it should be treated as a perturbative correction to the equation of motion in the same way as the non-linear terms.  
Therefore, it leads to an additional term to the list of non-linear terms in eq.~\eqref{NL terms},
\begin{equation}
\left\{ X Y,\quad 
\partial_i X\frac{\partial_i}{\partial^2}Y,\quad 
\frac{\partial_i\partial_j}{\partial^2} X\frac{\partial_i\partial_j}{\partial^2} Y,\quad
R_*^2\partial^2 X
\right\},
\end{equation}
where we keep $R_*^2$ as a bookkeeping parameter indicating $R_*^2k^2=\mathcal{O}(\del^2)$
with $k$ being the wavenumber of the Fourier mode in interest.
This leading derivative term would be activated only at the third iteration of eq.~\eqref{NL terms}, 
giving the nontrivial contributions at the fourth-order, which is out of the scope of this paper where given 
that we are limiting our analysis to the third order expansion in the biased tracer fields.
Contrary to the case of stochastic field, the higher derivative operators are allowed to correlate with 
any of the other deterministic operators. 
Note that any higher derivative operators including $(R^2_*\partial^2)^n\del$ 
can then be added in an analogous way by allowing them to contribute in the construction
of the basis via eq.~\eqref{NL terms} at the appropriate field order. 

Once Step 1 (subsec.~\ref{step_1}) and Step 2 (subsec.~\ref{step_2}) are completed, 
giving rise to the extended Monkey basis, Step 3 (subsec.~\ref{Step 3: Relate the coefficients}) 
proceed in the same way as done before, giving additional constraints for the new bias parameters. 

%====================================================================================%
\section{Comparison to Previous Works}
\label{sec:comp}
%====================================================================================%

In this section, we compare our result with three previous works; 
(i) McDonald \& Roy \cite{McDonald+:2009}, (ii) Mirbabayi, Schmidt \& Zaldarriaga \cite{Mirbabayi++:2014} (iii) EFT of LSS \cite{Senatore:2014}. 
There are many other papers featuring comparable bias expansion (see, e.g. \cite{Chan++:2012, Saito++:2014, Assassi++:2014, Eggemeier:2018}), 
but given that these should all be equivalent (at least up to the third-order) to the frameworks chosen above, we restrict our comparative study to these three. 
Note that in these previous formalisms, one needs to know the  non-linear dynamics of matter fluctuation to derive 
bias expansions, and their results were presented with the use of the SPT in the EdS Universe, whereas our monkey 
framework relaxes this assumption. Hence, $\mathcal{C}_{b}=1$ and $\mathcal{C}_{d}=3/14$ are implicitly adopted 
in these previous works, as we shall see below. All these previous works have seven free bias parameters 
and corresponding bias operator basis, making them equivalent to our monkey framework. 

%====================================================================================%
\subsection{McDonald\&Roy(2009)}
%====================================================================================%

McDonald and Roy  \cite{McDonald+:2009}, have found their basis expansion up to the third-order by fully exploiting the 
SPT result for the matter fluctuation $\delta_m$. They introduced the following bias operators 
\begin{equation}
\psi\equiv \eta-\frac{2}{7}s^2 +\frac{4}{21}\delta^2_m,
~~ s_{ij}\equiv \left(\frac{\partial_i\partial_j}{\partial^2}-\frac{1}{3}\delta_{ij}\right)\delta_m,
~~ t_{ij}\equiv \left(\frac{\partial_i\partial_j}{\partial^2}-\frac{1}{3}\delta_{ij}\right)\eta,
~~ \eta\equiv \theta_m -\delta_m.
\end{equation}
Then, for instance, they chose $\psi$ as an independent operator at the third order, because the SPT in the EdS 
universe shows $\eta=2s^2/7-4\delta^2_m/21$ up to the second order and  $\psi$ becomes non-zero from the third order. 
Through such arguments, the expansion for the bias trace field up to the third-order is obtained
\begin{equation}
\delta_h^{\rm MR} = c_\delta \delta_m + c_{\delta^2} \delta^2_m+c_{s^2} s^2
+c_{\delta^3}\delta^3_m+c_{\delta s^2}\delta_m s^2 + c_{st} st + c_{s^3} s^3+c_\psi \psi,
\label{MR expression}
\end{equation}
where $s^2\equiv s_{ij}s_{ij}, s^3\equiv s_{ij}s_{jl}s_{li}$ and $st\equiv s_{ij}t_{ij}$.
In order to show that this basis is, order by order, equivalent to the one obtained in the 
Monkey framework we compare it to the eq.~\eqref{monkey_bias} (using $\mathcal{C}_{b}=1$ and $\mathcal{C}_{d}=3/14$). 
Comparing the two bias expansion basis we see that $\delta_h^{\rm MR}$ can be re-expressed in 
terms of the Monkey framework by using the simple bias coefficient redefinition 
\begin{align}
\label{Monkey MR}
&a_1=c_\delta,\quad b_1=\frac{5}{7}c_\delta+c_{\delta^2}-\frac{1}{3}c_{s^2},\quad
b_2=c_\delta,\quad 
b_3=\frac{2}{7}c_\delta+c_{s^2},\\
&c_1=\frac{4}{9}c_\delta+\frac{10}{7}c_{\delta^2}-\frac{10}{21}c_{s^2}+c_{\delta^3}
-\frac{1}{3}c_{\delta s^2}+\frac{68}{441}c_{\psi}+\frac{2}{21}c_{st}+\frac{2}{9}c_{s^3},
\notag\\
&c_2=\frac{10}{7}c_\delta+2c_{\delta^2}-\frac{2}{3}c_{s^2},\quad 
c_3=\frac{1}{3}c_\delta+\frac{4}{7}c_{\delta^2}-\frac{4}{21}c_{s^2}+c_{\delta s^2}-\frac{4}{147}c_\psi-\frac{2}{21}c_{st}-c_{s^3},
\notag\\
&d_1=\frac{3}{14}c_\delta,\quad d_2=\frac{1}{2}c_{\delta},\quad d_3=\frac{2}{7}c_\delta,
\quad e_2=\frac{1}{2}c_\delta,\quad 
e_3=\frac{11}{63}c_\delta+\frac{4}{63}c_\psi-\frac{1}{2}c_{s^3},
\notag\\
&f_1=\frac{2}{21}c_\delta+\frac{10}{7}c_{s^2}-\frac{32}{147}c_\psi-\frac{2}{7}c_{st},\quad
f_2=\frac{2}{9}c_\delta+2c_{s^2}-\frac{8}{63}c_\psi+c_{s^3},
\notag\\
&f_3=\frac{8}{63}c_\delta+\frac{4}{7}c_{s^2}+\frac{40}{441}c_\psi+\frac{2}{7}c_{st}
+c_{s^3}.\notag
\end{align}
It is straightforward to confirm that the above expressions satisfy all the conditions on the monkey coefficients, 
$a_1, b_1,..,f_3$, derived in the previous section irrespective of McDonald and Roy's coefficients, $c_\delta, c_{\delta^2},...,c_{s^3}$.
Therefore, this bias expansion is a particular example of the Monkey formalism under the additional assumption of the SPT and the EdS universe. 
As is known, five bias parameters at third order in eq.~\eqref{MR expression}
implies that one third order operator (and its coefficient) out of five can be eliminated as a degenerate operator.

%====================================================================================%
\subsection{Mirbabayi, Schmidt \& Zaldarriaga(2014)}
%====================================================================================%

The authors of \cite{Mirbabayi++:2014} adopted the second spacial derivative of gravitational potential $\Phi$ as a fundamental building block for bias expansion,
\begin{equation}
\Pi_{ij}^{[1]}\equiv \frac{2}{3\Omega_m\mathcal{H}^2} \partial_i \partial_j \Phi
= \frac{1}{3}\delta^{\rm K}_{ij} \delta_m +s_{ij},
\end{equation}
where Poisson equation is used and an operator $\Pi^{[n]}_{ij}$ includes $n$-th and higher order contributions in perturbation. 
Contrary to the somewhat heuristic argument of McDonald and Roy, they provided a systematic scheme to find higher order bias terms from $\Pi_{ij}^{[1]}$. 
Since the time dependence of a $n$-th order operator is always $D^n(\tau)$ in the SPT\ in the EdS-like universe, 
taking its logarithmic derivative w.r.t. the growth factor $\dd/\dd\ln D=(\mathcal{H}f)^{-1}\dd/\dd\tau$ and multiplying  it by $n$ are equivalent at the leading order.
Thus their difference is $(n+1)$-th order and this fact leads to the following generating rule:
\begin{equation}
\Pi_{ij}^{[n+1]}=\frac{1}{n!}\left[
(\mathcal{H}f)^{-1}\frac{\rm D}{{\rm D} \tau}-
n\right]\Pi_{ij}^{[n]},
\end{equation}
where the convective (or Lagrangian) time derivative ${\rm D}/{\rm D}\tau\equiv \partial/\partial_\tau+v^i\partial/\partial x^i$ is also introduced to take into 
account the past trajectory $\bm x_{\rm fl}$ (see sec.~\ref{sec:canon}). The bias basis, at a given order, is then constructed out of
all possible scalar quantities made from $\Pi_{ij}^{[n]}$. In terms of  operators of McDonald and Roy basis it 
can be re-expressed as
\begin{align}
\label{MSZ operators}
{\rm 1st\ order:}\quad&{\rm Tr}[\Pi^{[1]}]=\delta_m,\\
{\rm 2nd\ order:}\quad&({\rm Tr}[\Pi^{[1]}])^2=\delta_m^2,\quad
{\rm Tr}[(\Pi^{[1]})^2]=s^2+\frac{1}{3}\delta_m^2,\notag\\
{\rm 3rd\ order:}\quad&({\rm Tr}[\Pi^{[1]}])^3=\delta_m^3,\quad
{\rm Tr}[(\Pi^{[1]})^2]{\rm Tr}[(\Pi^{[1]})]=\frac{1}{3}\delta_m^3+\delta_m s^2,\notag\\
&{\rm Tr}[(\Pi^{[1]})^3]=\frac{1}{9}\delta_m^3+\delta_m s^2+s^3,
\quad {\rm Tr}[\Pi^{[1]}\Pi^{[2]}]=\frac{17}{63}\delta_m^3+
\frac{16}{21}\delta_m s^2+s^3-\frac{5}{2}st, \notag
\end{align}
Thus MSZ bias expansion can again, up to third order, be expressed as another linear combination of operators in eq.~\eqref{MR expression}
(while $\psi$ is eliminated due to the degeneracy). One can find the corresponding monkey coefficient by remapping eq.~\eqref{Monkey MR} 
in accordance with eq.~\eqref{MSZ operators}.

%====================================================================================%
\subsection{EFT of LSS}
%====================================================================================%

The bias expansion of the EFT of LSS \cite{Senatore:2014, Angulo++:2015, Fujita++:2016} was derived by adding 
a physical component of ``non-locality in time'' to McDonald and Roy's operator basis $O_{\rm MR}$. 
The fact that biased tracers could be affected by the physics that happened  
on their past trajectory can be taken into account by dressing the operators in time integrals with time-dependent kernels.  We thus have
\begin{equation}
\delta_h^{\rm EFT}(t,\bm x) = \sum_{O_i \in O_{\rm MR}} \int^t dt' H(t')\,  c_{i}(t,t') \, O_i(t',\bm x_{\rm fl}),
\label{EFT operator}
\end{equation}
where $\bm x_{\rm fl}(t,t',\bm x) \equiv \bm x - \int_{t'}^{t} dt''\bm v(t'',\bm  x_{\rm fl}(t,t'',\bm x))$ is the past trajectory 
of biased tracer $\delta_h(t,\bm x)$ along which an operator $O_i$ has affected the formation of $\delta_h$ with unknown weighting factor $c_i$.
By perturbatively expanding the above expression, one apparently obtains much more bias operators than McDonald and Roy.
However, after resolving degeneracies between  operators, it turned out that the bias expansion is isomorphic (i.e., equivalent) to McDonald and Roy's one.%
\footnote{In the first \cite{Senatore:2014} and second paper \cite{Angulo++:2015}, new operators apparently appeared because of errors in calculations 
which were fixed in the third paper \cite{Fujita++:2016}.}
Therefore, it has been shown that non-locality in time eq.~\eqref{EFT operator} does not bring in any new bias operator up to third order in the EdS universe,
compared to other approaches.
In this context, our formalism clarifies that non-locality in time  never introduces new operators in any background universe (at least up to third order), 
given that the monkey theory does not acquire a new operator unless new non-linear terms are added to eq.~\eqref{NL terms} or the consistency relations are broken.

The bias expansion of the EFT of LSS is given by
\begin{align}
\delta_h = \tilde{c}_{\delta,1} \Big( \mathbb{C}^{(1)}_{\delta,1}+\mathbb{C}^{(2)}_{\delta,1}+\mathbb{C}^{(3)}_{\delta,1} \Big) &+ \tilde{c}_{\delta,2} \Big(\mathbb{C}^{(2)}_{\delta,2} +\mathbb{C}^{(3)}_{\delta,2}\Big)
+ \tilde{c}_{\delta^2,1}\Big( \mathbb{C}^{(2)}_{\delta^2,1}+\mathbb{C}^{(3)}_{\delta^2,1}\Big) \\
&+ \tilde{c}_{\delta,3} \mathbb{C}^{(3)}_{\delta,3}
+ \tilde{c}_{\delta^2,2} \mathbb{C}^{(3)}_{\delta^2,2} 
+ \tilde{c}_{s^2,2} \mathbb{C}^{(3)}_{s^2,2}
+ \tilde{c}_{\delta^3,1} \mathbb{C}^{(3)}_{\delta^3,1}, \notag
\end{align}
where $\tilde c_X$ are free bias parameters and $\mathbb{C}_X^{(n)}$ are
operators at $n$-th order whose definitions can be found in \cite{Angulo++:2015, Fujita++:2016}.
We again suppressed stochastic, higher derivative and counter terms in the above equation. The mapping of the coefficients is given by
\begin{align}
&a_1=\tilde c_{\delta,1},\quad b_1=\frac{5}{7}\tilde c_{\delta,2}+\tilde c_{\delta^2,1},\quad b_2=\tilde c_{\delta,1},\quad b_3=\frac{2}{7}\tilde c_{\delta,2}, \\
&c_1=\frac{4}{9}\tilde c_{\delta,3}+\frac{10}{7}\tilde c_{\delta^2,1}+
\tilde c_{\delta^3,1}-\frac{10}{21}\tilde c_{s^2,2},\quad
c_2=\frac{10}{7}\tilde c_{\delta,2}+2\tilde c_{\delta^2,1},\quad c_3=\frac{1}{3}\tilde c_{\delta,3}+\frac{4}{7}\tilde c_{\delta^2,2}-\frac{4}{21}\tilde c_{s^2,2},\notag\\
&d_1=\frac{3}{14}\tilde c_{\delta,1},\quad d_2=\frac{1}{2}\tilde c_{\delta,1}
\quad d_3=\frac{2}{7}\tilde c_{\delta,1},\quad e_2=\frac{1}{2}\tilde c_{\delta,1},
\quad e_3=\frac{2}{7}\tilde c_{\delta,2}-\frac{1}{9}\tilde c_{\delta,3}-\tilde c_{s^2,2},\notag\\
&f_1=\frac{2}{21}\tilde c_{\delta,3}+\frac{10}{7}\tilde c_{s^2,2},
\quad f_2=\frac{2}{9}\tilde c_{\delta,3}+2\tilde c_{s^2,2},\quad
f_3=\frac{8}{63}\tilde c_{\delta,3}+\frac{4}{7}\tilde c_{s^2,2}. \notag
\end{align}
It can be easily confirmed that all the conditions on the monkey coefficients are always satisfied for arbitrary EFT bias parameters $\tilde c_X$.

%====================================================================================%
\section{Summary and Discussion}
\label{sec:summa}

Both Eulerian and Lagrangian biasing schemes employ the two-step procedure wherein bias expansion is introduced in the one-step, and 
perturbation theory is used to describe dark matter dynamics in the second step. 
This requires the computational tasks where order by order similar perturbative expansions are done repeatedly, 
increasing the complexity of the description in either scheme. Moreover, in both schemes, the former step relies on 
a somewhat arbitrary way of how the bias expansion basis is chosen at each perturbative order. 

In this paper, we have developed the new formalism for constructing the biased tracer field (such as galaxies, 
21cm lines, clusters, etc.), which we dub the Monkey bias theory. The motivation for such an approach was to avoid the 
redundancy in the basic construction  present in all other approaches and to provide a 
self-consistent basis construction algorithm based on physical principles. We summarize below several key features of this approach. 

\begin{itemize}

\item Monkey bias expansion is a direct expansion of the biased tracer field, which originally includes all possible 
bias terms at given perturbative order and then gets constrained into the physical basis by consistency relation of large scale 
structure derived from the equivalence principle and adiabatic perturbations. 

\item In addition to the direct utilization of consistency relations, 
additional spurious dependence on the direction of soft modes remains that is not killed by the consistency conditions alone. 
Disregarding such contributions, in principle, constitutes an additional physical requirement, separate from the current form of consistency relations, and thus prompts further investigation that shall be addressed in future work. 

\item These physical constraints are imposed on the tree-level statistics 
where the computational effort is minimal, which automatically guarantees the loop contributions are also physical. 
Thus our formalism ensures the completeness and IR safety of the generated bias basis. As a consequence of this 
construction method, we also do not need to worry if an operator is missed in our bias expansion.

\item Construction of the Monkey bias basis for a given perturbative order is formulated as a three-step algorithmic procedure that yields 
itself to automatized computer implementation. It can be especially useful to construct the basis at 
higher orders in perturbation theory where a fully automated approach would guarantee consistency and completeness 
in the ``no operator left behind'' sense. This feature, in particular, we see as an advantage compared to all the current 
bias tracer frameworks, where the construction of the higher orders can become a cumbersome endeavor. 

\item Applying this construction procedure, we obtain explicit results of the bias tracer field up to the third order in the 
perturbative expansion (counting by numbers of linear fields $\del$ involved) as given in eq.~\eqref{monkey_bias}. 
If the EdS universe is additionally assumed, our Monkey bias expansion is equivalent to the ones in the literature. 
We explicitly compare the results in \cite{McDonald+:2009,Senatore:2014, Mirbabayi++:2014} 
to ours, and give explicit linear mappings of bias coefficients. 

\item For the first time, we include the exact time evolution effects in the generalized background universe beyond the EdS universe in the bias expansion.
These emerge as the universal coefficients, $\mathcal{C}_b$ and $\mathcal{C}_d$, in our main result eq.~\eqref{monkey_bias}.
Even though their deviation from the standard EdS-like evolution is not likely to be dramatic, they are not degenerate with the bias parameters, 
contrary to the standard lore. This is because the coefficients of the bulk displacements operators are protected by the equivalence principle, 
but at third-order the bias expansion can exhibit the nonlinear displacement effects whose coefficients depends on the background evolution 
in the $\Lambda$CDM universe. These time evolution effects can appear even at the second-order and become diversified in the beyond 
$\Lambda$CDM theories (see, e.g., \cite{Sefusatti+:2011,Anselmi++:2011, Fasiello+:2016a,  Lewandowski:2019}).
Isolating this contribution can serve as a clean test of $\Lambda$CDM dynamics. 
As a simplest observable of these evolution effects, we suggest the near-optimal bispectrum estimators for the displacement contributions as is proposed 
in, e.g., \cite{Schmittfull++:2014} for other bias coefficients.

\end{itemize}

In the paper, we focused primarily on obtaining the results for so-called deterministic bias operators. 
However, we also discuss the extension of the Monkey framework to include the stochastic bias contributions, 
as well as higher derivative operators. Both of these extensions can be relatively easily added to the framework, 
modifying just the first step of the proposed algorithm, while the application of the consistency conditions remains 
unchanged. 

As a potential benefit and application of the approach, we envisage it in a more straightforward construction of 
the higher-order contributions such as one-loop bispectrum or two-loop power spectrum of biased tracers, 
as well as in the determining and exploiting the approximate degeneracies and marginal operators.
The latter is of particular interest given that the number of free parameters in constructed observables can become 
quite large, and thus, it is of interest to reduce the number of degrees of freedom to the relevant ones.
Furthermore, the extension of the formalism to incorporate the redshift space distortions, as well as the expansion of
bias tensor fields (see e.g. \cite{Vlah++:2019} for a recent treatment) should be of interest and charts some of the 
future developments.
Finally, we note that, in our opinion, presented Monkey formalism for the perturbative construction of the generic 
bias tracer fields utilizes one of the simplest and yet most explicit applications of the consistency relations of large scale structure so far. 

%====================================================================================%
\acknowledgments

We would like to thank Paolo Creminelli, Fabian Schmidt, Sergey Sibiryakov and Marko Simonovi\' c for the useful discussions 
and helpful comments. The work of TF was supported by JSPS KAKENHI No. 17J09103 and No. 18K13537.
No animals were harmed during any stage of this project. 

\appendix

%====================================================================================%
\section{SPT result in the generalized $\Lambda$CDM universe}
\label{app:lcdm_bias}
%====================================================================================%

In this section, we show that Monkey basis given in eq.~\eqref{deltah list} is sufficient to capture the 
exact time dependence of the perturbative dark matter solutions in the $\Lambda$CDM universe and its simple quintessence extensions. 

The usual practice in the field of perturbative LSS is to rely on so-called EdS-like approximation where the perturbative kernels ($F_n$ and $G_n$ in SPT) 
are assumed to be time-independent, and each perturbative order is scaled by the linear growth rate $D(\tau)^n$. 
For the $\Lambda$CDM universe, this turns out to be a good approximation yielding a $\sim1\%$ 
accuracy \cite{Takahashi:2008, Fasiello+:2016a, Fasiello+:2016b}.
However, one can exactly solve the SPT in the $\Lambda$CDM universe 
with the time-dependent perturbative kernels (see, e.g., \cite{Fasiello+:2016a}). 
The exact solution for the density fluctuation $\delta_m$ 
in the extended $\Lambda$CDM universe can be written in our bias expansion basis as 
\begin{align}
\label{SPT LCDM}
&a_1^{(m)}=1,\ b_1^{(m)}=\frac{1}{4}\left(3\nu_2-2+2\epsilon_1\right),\  b_2^{(m)}=1-\epsilon_1,\  b_3^{(m)}=\frac{3}{2}\left(1-\frac{1}{2}\nu_2-\epsilon_1\right),\,
\\
&c_1^{(m)}=\frac{1}{8} (-9 \lambda_1+12 \lambda_2+3 \nu_3-6 \nu_2 +2+6\nu_2\epsilon_1-2\epsilon_2),\
c_2^{(m)}=\frac{1}{2}\left(3 \nu_2-2-3\nu_2\epsilon_1+2\epsilon_2\right),\notag\\ &c_3^{(m)}=-\frac{3}{8} (-13 \lambda_1+12 \lambda_2+3 \nu_3-6 \nu_2 +2+6\nu_2\epsilon_1-2\epsilon_2),\,
\notag\\ 
&d_1^{(m)}=\frac{1}{8}  (3 \lambda_1-12 \lambda_2+3 \nu_3-2+2\epsilon_2),\  d_2^{(m)}=\frac{1}{2}(1-\epsilon_2),\  d_3^{(m)}=-\frac{3}{8} (\lambda_1-4 \lambda_2+\nu_3-2+2\epsilon_2),\notag\\
&e_2^{(m)}=\frac{1}{2}(1-\epsilon_2),\, e_3^{(m)}=-\frac{3}{8}(-5\lambda_1+4\lambda_2+\nu_3-2+2\epsilon_2),
\notag\\ 
&f_1^{(m)}=-\frac{3}{8}  (\lambda_1+4 \lambda_2-\nu_3-2 \nu_2 +2+2\nu_2\epsilon_1-2\epsilon_2),\notag\\
&f_2^{(m)}=\frac{3}{4}  (-5 \lambda_1+4 \lambda_2+\nu_3-2 \nu_2+2+2\nu_2\epsilon_1-2\epsilon_2),\notag\\
&f_3^{(m)}=\frac{3}{8} (-9 \lambda_1+12 \lambda_2+\nu_3-6 \nu_2+6+6\nu_2\epsilon_1-6\epsilon_2), \notag
\end{align}
where $\nu_2(\tau),~\nu_3(\tau),~\lambda_1(\tau), ~\lambda_2(\tau)$ represent the influence of the cosmic expansion on the structure formation, 
and $\epsilon_1(\tau), ~\epsilon_2(\tau)$ encode the effect of quintessence fluctuation which satisfy $(1-\epsilon_1)^2=1-\epsilon_2$ (see e.g. \cite{Sefusatti+:2011, Fasiello+:2016b}).
In the limit of the standard $\Lambda$CDM universe, $\epsilon_1$ and $\epsilon_2$ vanish.
One can explicitly check that all the conditions found in subsec.~\ref{Step 3: Relate the coefficients} hold irrespective of these functions of time.
It can be also confirmed that the above results are reduced into eq.~\eqref{SPT EdS} in the EdS limit.

%====================================================================================%
\section{One-loop Power Spectrum}
\label{app:loop_ps}
%====================================================================================%

In Section~\ref{sec:monkey} we have obtained the explicit field expansion up to the third order in the Monkey basis.
Besides the tree-level statistics up to the trispectrum, this also allows us to compute the one-loop power spectrum
with soft loop momentum. The consistency relations show that the leading $\mathcal{O}(q^{-2})$
term in the integrand in the unequal-time correlator should vanish in the equal-time limit \cite{Creminelli++:2013b},
\begin{align}
\label{soft-loop_cc}
\langle\delta_{h_\alpha}(\bm k,\tau_1) & \delta_{h_\beta}(\bm -\bm k,\tau_2) \rangle_{\rm 1-softloop}' \\
&\approx -\frac{1}{2}\langle\delta_{h_\alpha}(\bm k,\tau_1) \delta_{h_\beta}(\bm -\bm k,\tau_2) \rangle_{\rm tree}'\int^\Lambda
\frac{\dd^3 q}{(2\pi)^3} \big[D(\tau_1)-D(\tau_2)\big]^2\left(\frac{\bm q\cdot \bm k}{q^2} \right)^2
P_\ell(q), \notag
\end{align}
where $\Lambda$ denotes a UV cutoff which restricts the loop momentum to be much less than the external one, $q\le \Lambda\ll k$.
The cross-correlation one-loop power spectrum for biased tracers $h_\alpha$ and $h_\beta$ can be written as
\begin{align}
P_{h_\alpha h_\beta}^{\rm 1loop}(k)&= 2\int_{\bm q}K_2^{(\alpha)}(\bm{k-q},\bm{q})
K_2^{(\beta)}(\bm{k-q},\bm{q})P_\ell(q)P_\ell (|\bm k-\bm q|) \\
&\hspace{2cm}+3P_\ell(k)\int_{\bm q}\left[a_1^{(\alpha)} K_3^{(\beta)}(\bm{k},\bm{ -q},\bm{q})+
a_1^{(\beta)}K_3^{(\alpha)}(\bm{k},\bm{ -q},\bm{q})\right]P_\ell(q), \notag
\label{bare Phh}
\end{align}
where $\int_{\bm q}\equiv\int \dd^3 q/ (2\pi)^3$, and $K_2$ and $K_3$ include the contributions from 
the second and third order operators whose expressions are written in eq.~\eqref{K_kernels}. 
Before taking the soft limit, we consider the limit when the two momenta in the $K_3^{(h)}$ kernel 
are the of the same magnitude and direction, but of the opposite orientation. 
We thus consider a limit $\bm k_1 \to \bm k$, $\bm k_2 \to \bm - \bm q$ and $\bm k_3 \to \bm q$
using a probe $\bm \epsilon$ field of vanishing magnitude. We have
\begin{equation}
K_3^{(h)}(\bm k,\bm \epsilon-\bm q,\bm q)
= \left(d_1^{(h)}-d_2^{(h)}+d_3^{(h)}\right) \frac{\bm k\cdot \bm \epsilon}{3\epsilon^2}
+\left(f_1^{(h)}-f_2^{(h)}+f_3^{(h)}\right) \frac{(\bm k\cdot \bm \epsilon)^2}{3k^{2}\epsilon^2}
+\cdots,
\end{equation}
The field $\bm \epsilon$ is thus introduced as a regulator and the kernel should be free of divergencies and 
orientation dependence of $\bm \epsilon$, in the limit $\epsilon\to 0$. This imposes the two constraints
\begin{equation}
d_1^{(h)}-d_2^{(h)}+d_3^{(h)}=0,
\qquad
f_1^{(h)}-f_2^{(h)}+f_3^{(h)}=0,
\label{1loop f123}
\end{equation}
which are already satisfied by the constraints obtained from the trispectrum consideration in eqs.~\eqref{single IR},~\eqref{double IR}, and \eqref{f123}.
This is, of course, expected given that the $K_3^{(h)}$ kernel configuration entering the one-loop power spectrum is 
a simplification of the more general trispectrum case. 
Using the Monkey bias expansion in eq.~\eqref{deltah list}, 
the leading soft one-loop contribution is then given by
\begin{equation}
\left.P_{h_\alpha h_\beta}^{\rm 1loop}(k)\right|_{\rm softloop}\approx \left[b_{2}^{(\alpha)}b_{2}^{(\beta)} 
- a_1^{(\alpha)} e_2^{(\beta)}- a_1^{(\beta)} e_2^{(\alpha)}\right]\frac{1}{3}P_\ell(k)\int_{\bmq}^\Lambda \frac{k^2}{q^2}P_\ell (q).
\label{1loop divergence}
\end{equation}
The soft-loop consistency condition expressed in eq.~\eqref{soft-loop_cc} requires that the pre-factor above vanishes. 
However, this condition is the same as the trispectrum condition given in eq.~\eqref{soft two legs}.
Therefore, as was to be expected, the one-loop power spectrum does not require any new conditions and obtains
the required IR safe form by inheriting the constraints obtained from imposing the consistency condition on the three-level statistics.
The expectation is that the same trend, shown here at the level of the one-loop power spectrum, extends to the 
higher loops and higher-order statistics. For example, that the one-loop bispectrum and two-loop power spectrum 
are fully IR safe and regular, once the consistency conditions are imposed on the corresponding higher-order tree-level spectra.

%====================================================================================%
\section{Expressions for Observables}
\label{app:obs}
%====================================================================================%

In this section, we summarise expressions for the observable statistics used in the paper. 
Namely, we relay on the one-loop power spectrum, three-level bispectrum, and three-level trispectrum.
This constitutes all the statistics that are ``unlocked'' by considering the field expansion up to the third-order. 
The un-symmetrized kernels at the second-order $K_2(\bm{q}_1,\bm{q}_2)$ and the third-order 
$K_2(\bm{q}_1,\bm{q}_2,\bm{q}_3)$ can be obtained by Fourier transforming the final Monkey basis given in eq.~\eqref{monkey_bias}.
We have
\begin{align}
\label{K_kernels}
K_2&=\Blue{\mathcal{C}_b} \Red{a_1} \frac{\bmq_1\cdot \bmq_2}{q^2_2}+\Red{b_1}+\Red{b_3}\frac{(\bmq_1\cdot\bmq_2)^2}{q^2_1q^2_2}, \\
K_3&=\Blue{\mathcal{C}_{b}}   \frac{\bmq_2\cdot \bmq_3}{q^2_3}
\left[2 \Red{b_1}+\Red{b_3}\frac{\bmq_1\cdot(\bmq_2+\bmq_3)}{q_1^2}\frac{(\bmq_2\cdot\bmq_3)}{q^2_2}\right]
+\Blue{\mathcal{C}_{d}}\Red{a_1}\frac{\bmq_1\cdot(\bmq_2+\bmq_3)}{|\bmq_2+\bmq_3|^2}\left[1
-\frac{(\bmq_2\cdot\bmq_3)^2}{q^2_2q^2_3}\right]
\notag\\
&+\frac{1}{2}\Blue{\mathcal{C}_{b}^2} \Red{a_1}\frac{\bmq_2\cdot \bmq_3}{q^2_3} 
\left[\frac{\bmq_1\cdot(\bmq_2+\bmq_3)}{q_1^2}+ \frac{\bmq_1\cdot(\bmq_2+\bmq_3)}{|\bmq_2+\bmq_3|^2}\left\{1+\frac{(\bmq_2\cdot\bmq_3)}{q^2_2} \right\}\right]
 \notag\\
&+\Red{c_1} +\Red{c_3}\frac{(\bmq_2\cdot\bmq_3)^2}{q^2_2q^2_3} 
+\Red{f_1}\frac{\left[\bmq_1\cdot(\bmq_2+\bmq_3)\right]^2}{q_1^2|\bmq_2+\bmq_3|^2}\left[1
-\frac{(\bmq_2\cdot\bmq_3)^2}{q^2_2q^2_3}\right]
\notag\\
&+\Red{f_2}\frac{\bmq_2\cdot \bmq_3}{q^2_3}\left[\frac{\left[\bmq_1\cdot(\bmq_2+\bmq_3)\right]^2}{q_1^2|\bmq_2
+\bmq_3|^2}\left\{1+\frac{(\bmq_2\cdot\bmq_3)}{q^2_2}\right\}  - \frac{1}{2}\frac{\bmq_1\cdot(\bmq_2+\bmq_3)}{q_1^2}\frac{(\bmq_2\cdot\bmq_3)}{q^2_2}\right],
\notag
\end{align}
where we employ the same color scheme for the bias parameters and the universal coefficients as in eq.~\eqref{monkey_bias}.
For the reader's convenience, let us quote the characteristic universal coefficient values again. 
If the matter dynamics is described by the SPT-like equations (including its EFT extensions) 
in the $\Lambda$CDM universe, assuming the EdS-like approximation of the time dependence, 
we can use $\mathcal{C}_{b}=1$ and $\mathcal{C}_{d}=3/14$.

Once the kernel functions are symmetrized, we can write explicit expressions for the one-loop power spectrum, 
bispectrum, and trispectrum. The power spectrum is given by
\begin{align}
P_{h_\alpha h_\beta}^{\rm 1loop}(k)
&= a_1^{(\alpha)}a_1^{(\beta)} P_\ell(k) \\
&\hspace{0.5cm} + 2\int_{\bm q}K_2^{(\alpha)}(\bm{k-q},\bm{q})K_2^{(\beta)}(\bm{k-q},\bm{q})P_\ell(q)P_\ell (|\bm k-\bm q|) \notag \\
&\hspace{0.5cm} + 3P_\ell(k)\int_{\bm q}\left[a_1^{(\alpha)} K_3^{(\beta)}(\bm{k},\bm{ -q},\bm{q})+
a_1^{(\beta)}K_3^{(\alpha)}(\bm{k},\bm{ -q},\bm{q})\right]P_\ell(q). \notag
\end{align}
The power spectrum is the only statistics that exhibits the loop contributions at the field order we work at (third-order in the fields). 
The constraints obtained from the consistency conditions ensure that the power spectrum exhibits all the expected IR-safe 
behavior. 

Tree-level bispectrum is given by  
\begin{align}
B_{h_\alpha h_\beta h_\gamma}(\bmk_1,\bmk_2,\bmk_3)&= 2a_1^{(\alpha)}a_1^{(\beta)} K_2^{(\gamma)}(\bmk_1,\bmk_2)P_\ell(k_1)P_\ell(k_2) + 2\, {\rm perm},
\end{align}
where only second order fields contribute. 

Finally, the tree-level bispectrum is given by
\begin{align}
T_{h_{\alpha} h_\beta h_\gamma h_\delta}(\bmk_1,\bmk_2,\bmk_3,\bm k_4)&\\
&\hspace{-3cm}=6\Big[a_1^{(\alpha)}a_1^{(\beta)}a_1^{(\gamma)}K_3^{(\delta)}(\bm k_1,\bm k_2,\bm k_3)
P_\ell(k_1)P_\ell(k_2)P_\ell(k_3) +3\,{\rm perm} \Big] \notag\\
&\hspace{-1cm}  + 4\Big[a_1^{(\alpha)}a_1^{(\beta)} P_\ell(k_1)P_\ell(k_2) \Big\{P_\ell(|\bmk_2+\bmk_4|) 
K_2^{(\gamma)}(-\bmk_1, \bmk_1+\bmk_3) \notag\\
&\hspace{3cm} \times K_2^{(\delta)}(-\bmk_2, \bmk_2+\bmk_4)+(\bmk_1\leftrightarrow\bmk_2)\Big\} + 5\, {\rm perm}\Big], \notag
\end{align}
where again all three kernels contribute. 
%

%====================================================================================================%
\bibliography{ms}

\end{document}